# Effect of Transmission Line Switching on Damping of Interarea Modes

Vaithianathan "Mani" Venkatasubramanian and Menuka, Washington State University, Pullman WA

Ran Xu, California Independent System Operator, Folsom, CA

Ulyana Elliott and Gary Kobet, Tennessee Valley Authority, Chattanooga, TN

***Abstract***— This paper will explore the impact of switching transmission lines on the damping of interarea oscillation modes. This phenomenon played a central role in the 2025 Iberian blackout and in some North American oscillation events. The paper first assesses the effectiveness of line switching for enhancing mode damping by studying the problem in classical test power systems. The paper then tests these observations in realistic large-scale models of the Western Interconnection that represent recent real-time system conditions. It is shown that the switching of key transmission lines can have significant impact on the damping of inter-area modes depending on system operating conditions.

*Index Terms*—Oscillations, inter-area modes, operator actions, transmission lines.

## I. Introduction

System oscillations related to oscillatory modes occur in power systems frequently caused by small-signal and large disturbances. These oscillations remain mostly well-damped, though undamped, sustained, and growing oscillations can appear because of stressed operating conditions and improper control designs. A famous example is the August 10, 1996 Western American blackout that was caused by a negatively damped 0.25 Hz North-South inter-area mode [1]. During this event, the loss of several 500-kV transmission lines gradually reduced the damping of the inter-area mode [1], which was a root cause mechanism of the blackout. Subsequently, several transmission operators have instituted operating procedures and policies recommending the insertion of series capacitor banks and switching in transmission lines to mitigate problematic oscillations.

Recently, the grid operators in Spain switched in multiple transmission lines to dampen oscillations seen in the initial phases of the April 28, 2025 Iberian blackout event, which worsened the overvoltage conditions that eventually led to the blackout [2]. This paper first discusses these events to provide a historical background on oscillation events and operator and relay actions related to transmission lines. Then, we study inter-area mode damping and transmission system strength in the classical Kundur test power system [3] and a validated model of the August 10, 1996 western blackout [1]. These studies show that the effectiveness of transmission line switching on inter-area mode damping is sensitive to the overall strength of the system in terms of system conditions including critical tie-line flows. Finally, we test these relationships in archived real-time large-scale operational system models of the Western Interconnection, which shows that the damping improvement (or reduction) depends on the choice of the transmission lines being switched.

## II. System Oscillation Events

For motivation, we discuss three events in this section: 1) August 10, 1996 western system blackout, 2) April 28, 2025 Iberian blackout, and 3) April 6, 2026 eastern interconnection (EI) event.

### II.1 August 10, 1996 Western blackout event:

The blackout was caused by diverging inter-area oscillations associated with the North-South 0.25 Hz Western interconnection mode. The oscillations on the California-Oregon Intertie (COI) lines grew from about 25 MW to 1600 MW in about seventy seconds as shown in Fig.1. Solid lines show the time-plot of archived synchrophasor measurements, while dashed lines show the corresponding time-plot from a validated simulation model [1]. The validated large-scale western system model from [1] is used for damping sensitivity analysis later in Section III.

Fig. 2 shows the damping estimation using archived synchrophasor data from the event in [4], where the mode estimation was done using a stochastic subspace identification algorithm. The plot shows that the damping of the 0.25 Hz mode decreased from about 8% to less than 4% and then to 2% with the tripping of two 500-kV lines, namely from Allston to Keeler and Keeler to Pearl at around 7 seconds in Fig. 2. While the Keller-Allston line tripped because of the line sagging and touching a tree that caused a fault, the subsequent relay action inadvertently led to the tripping of the Keeler to Pearl line. The loss of these two lines reduced the damping of the 0.25 Hz mode by about 6% as shown in Fig. 2. The analysis of the historical event in [4] shows the significant impact transmission line switching can have on the damping of inter-area modes.

Even today, several Western transmission system operators have standing operating procedures to insert series capacitor banks and/or switching in transmission lines to improve damping of inter-area modes when problematic oscillations occur.

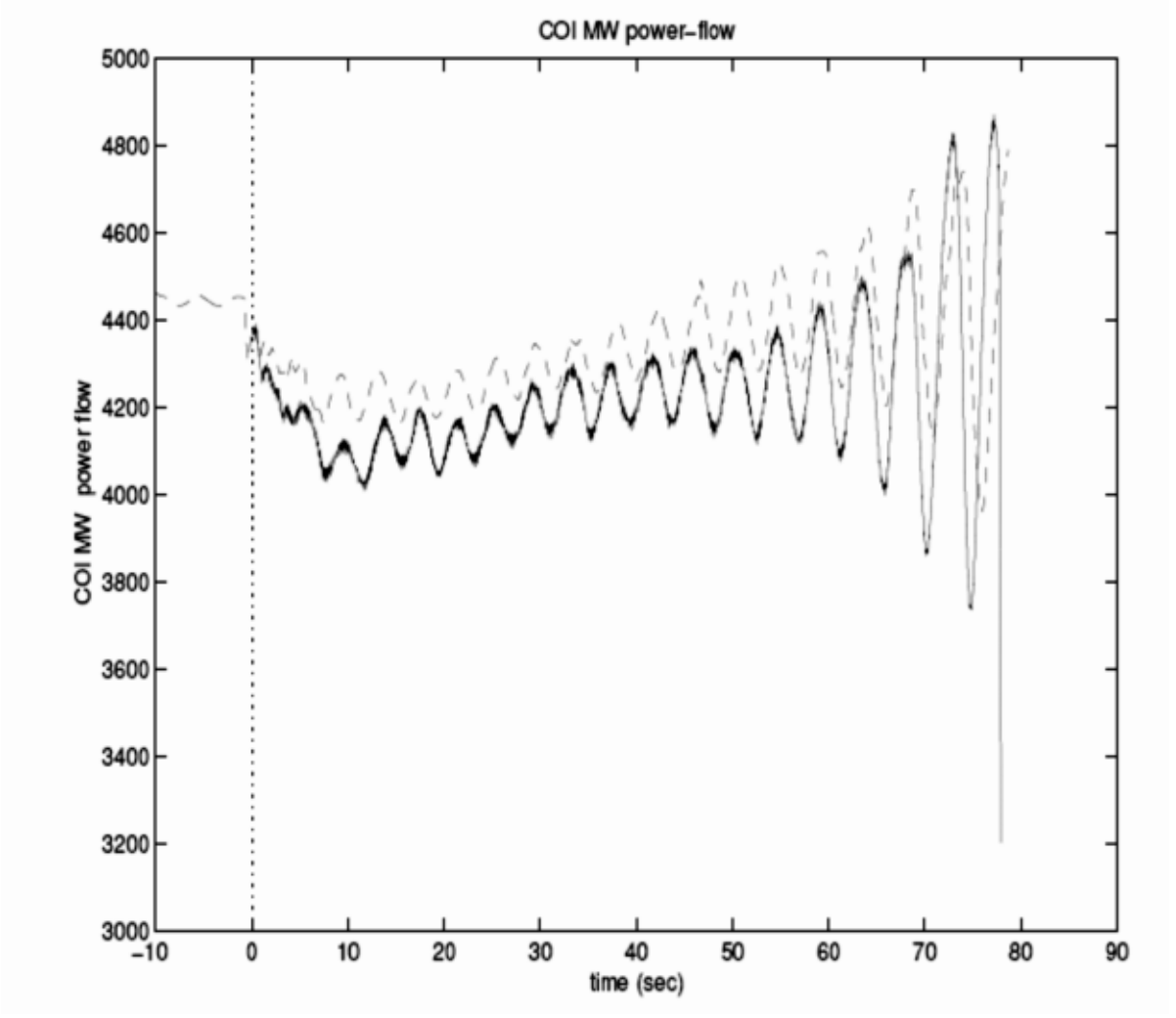

Figure 1: Growing oscillations on California-Oregon 500 kV AC transmission lines on August 10, 1996 [1].

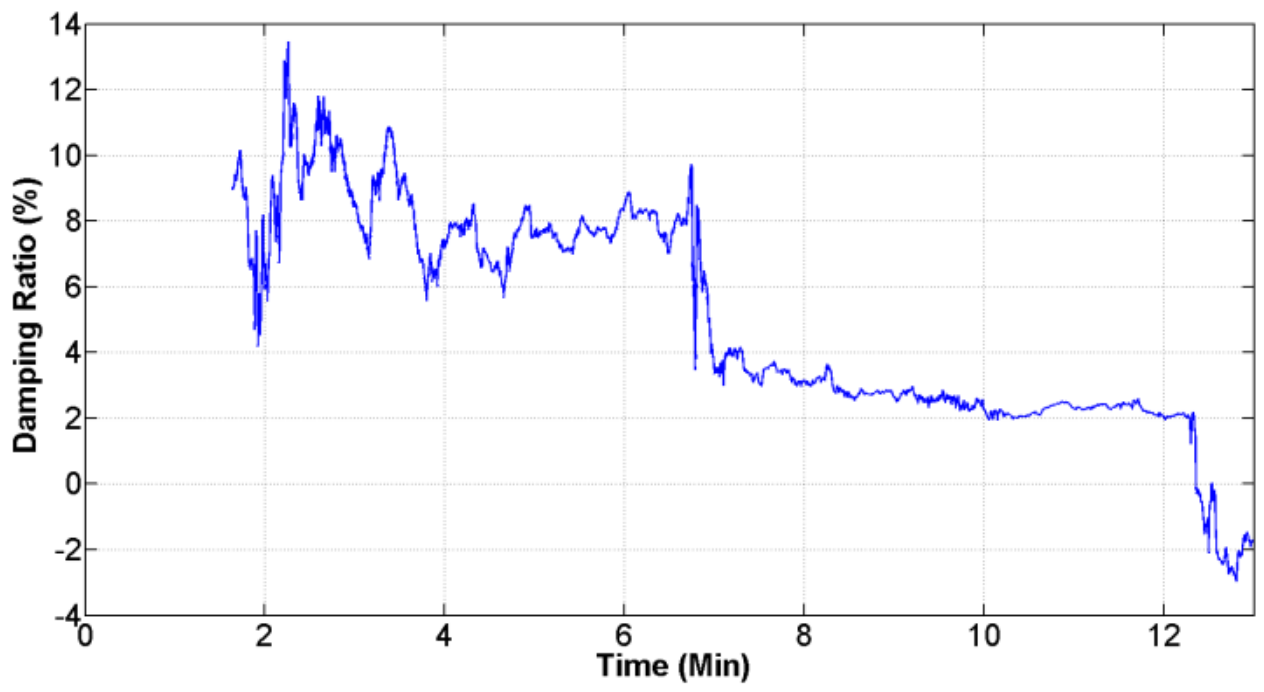

Figure 2: Evolution of the 0.25 Hz mode damping during the August 10, 1996 event [4].

II.2. April 28, 2025 Iberian blackout event:

The Iberian Peninsula, including the countries of Spain and Portugal, experienced a catastrophic blackout on April 28, 2025. Fig. 3 shows two large oscillations observed all over the European Interconnection ENSTOE before the blackout [2]. The first oscillation, which lasted between 12:03 and 12:07 pm, was at 0.63 Hz, while the second oscillation, which was between 12:19 and 12:21 pm, was at 0.2 Hz.

Synchrophasor data archived during the event was used to estimate and analyze the evolution of these two oscillatory modes in [5]. Fig. 4 shows the evolution of the 0.63 Hz mode, while Fig. 5 shows that of the 0.2 Hz mode. Specifically, the report [2] notes that operators in Spain took several switching actions during the two oscillation events to mitigate them. This included reducing MW power export from Spain to France and switching in 11 transmission lines in Spain to improve oscillation damping. The effects of the operator actions are reflected in the damping estimates shown in Figs. 4 and 5. In Fig. 4, the damping of the 0.63 Hz mode increases from near zero to about 6% around 12:08 pm, causing the oscillation to damp out correspondingly. Similarly, the damping of the 0.2 Hz mode increases from near zero to around 4% at around 12:21 pm, thus damping out the oscillation in the time plot. While switching 11 transmission lines helped improve damping of the 0.63 Hz and 0.2 Hz oscillations, it worsened the overvoltage conditions in Spain, which was a root cause of the Iberian blackout [2].

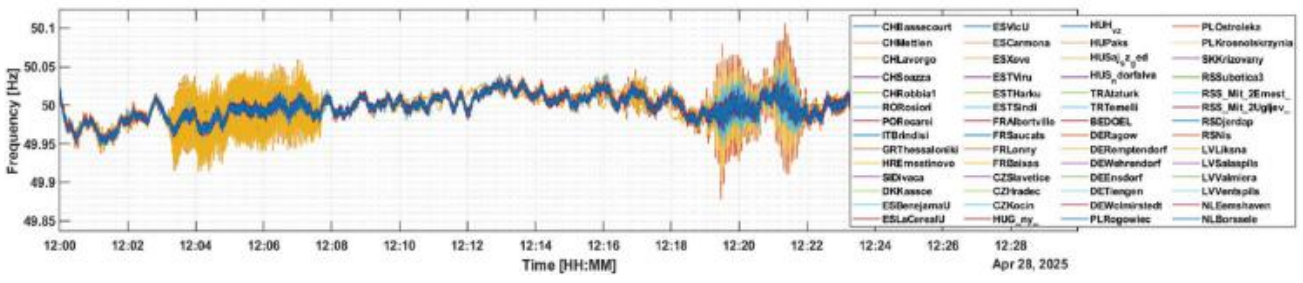

Figure 3: Two different oscillations observed during the April 28, 2025 Iberian event [2].

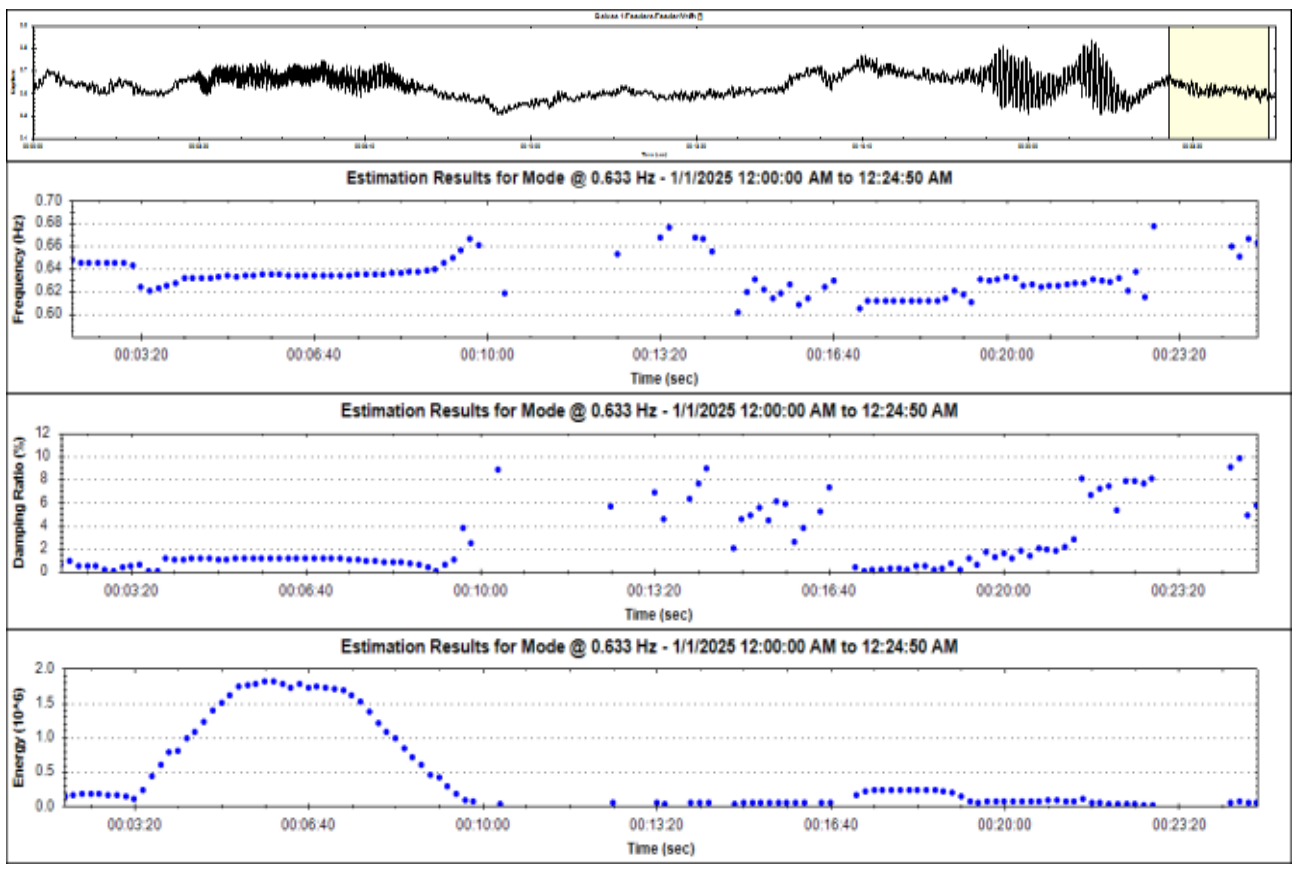

Figure 4: Evolution of the 0.63 Hz mode during the Iberian event [5]

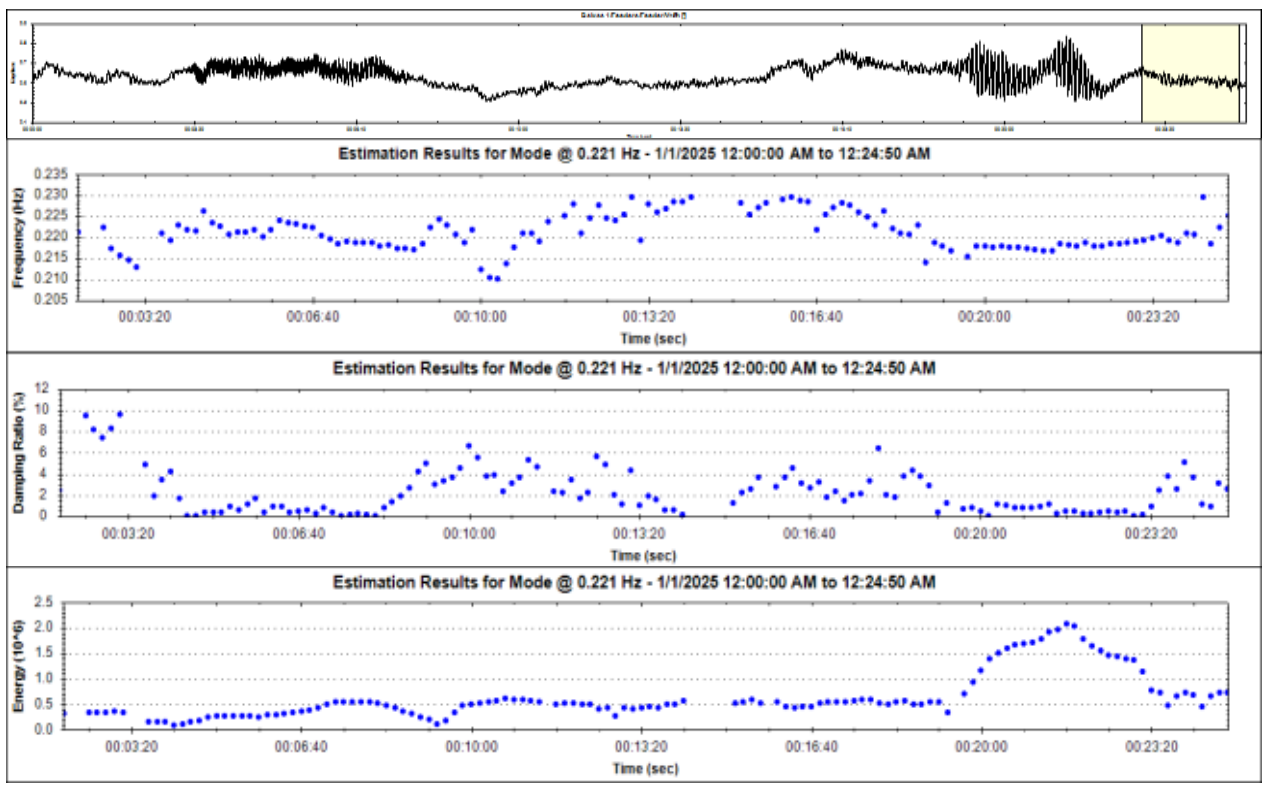

Figure 5: Evolution of the 0.2 Hz mode during the Iberian event [5]

Spanish operators' transmission-line switching during this event partially motivated this paper by inspiring us to examine this approach in more detail.

II.3. April 6, 2026 EI event:

The event started with the line connecting substations L and V in Fig. 6 already out of service. Then, the 500 kV line connecting substations C1 and C2 was disconnected to work on a 500 kV switch in substation C1. This region near substation C1 includes several PV power plants and became weak, with a short-circuit ratio (SCR) near 4, after the C1-C2 transmission line tripped. This led to the spontaneous emergence of MVAR (and voltage) oscillations, which were seen over a wide geographical area, as shown in Fig. 7. When the C1-C2 transmission line was switched back into service, the oscillations damped out. Initial analysis indicates that interactions between the solar plant controllers and the neighboring system may have become unstable when the C1-C2 transmission line was switched out, and subsequently became stable again when the line was switched in. The observed oscillations were at 0.22 Hz, which may have weakly resonated [6] with EI inter-area modes. In this paper, this event is another example of mode damping changing significantly when a critical transmission line is switched.

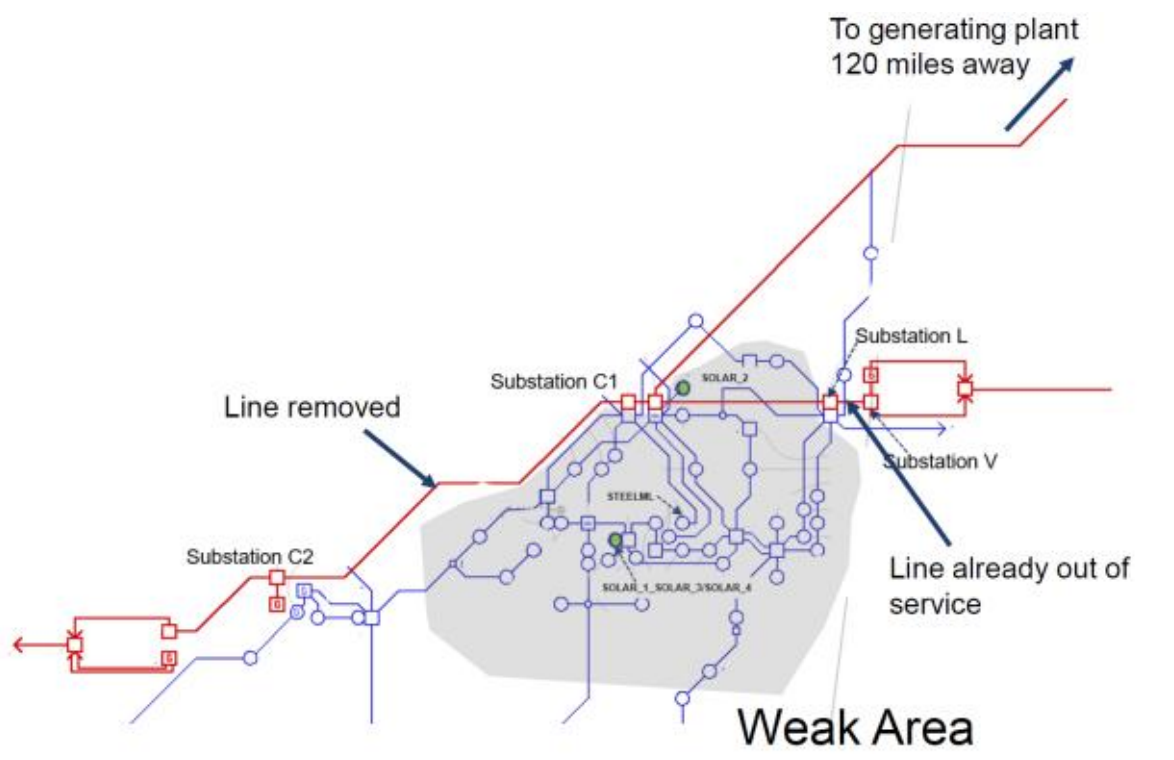

Figure 6: System conditions during the 2026 EI event.

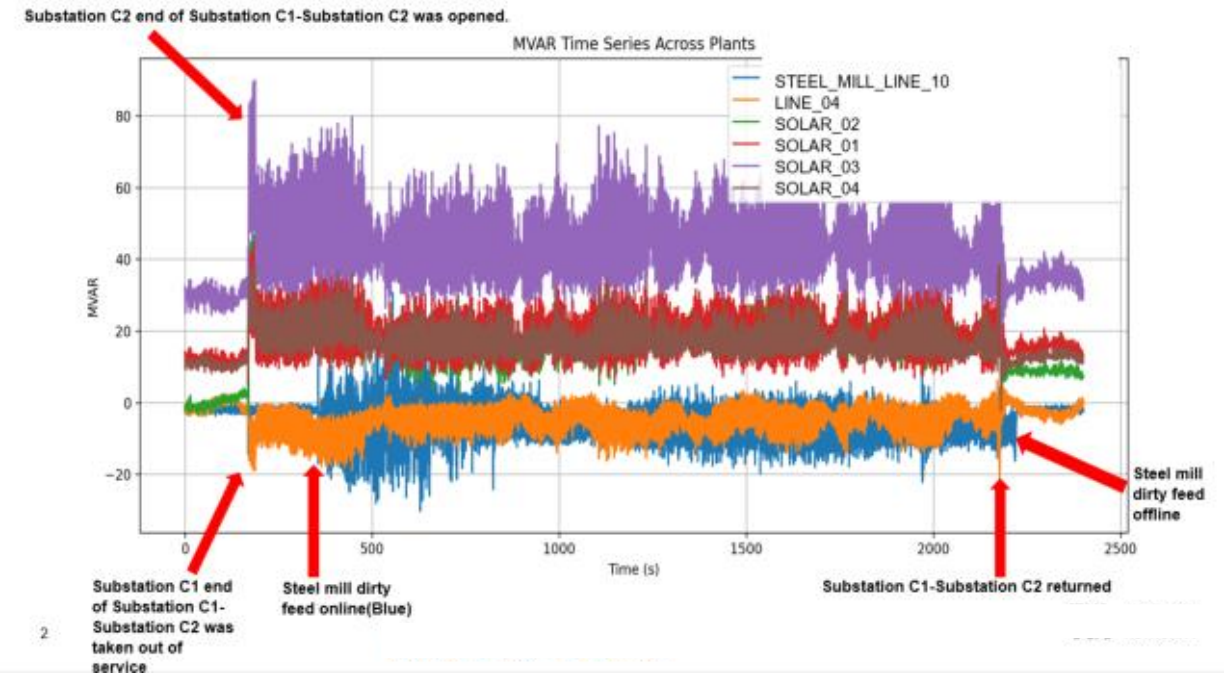

Figure 7: MVAR oscillations during the 2026 EI event.

## III. Power Grid Test Systems

We tested the observed phenomenon using both small- and large-scale test systems, as described in the following subsections. For the small-scale studies, we used three models: 1) a simplified version of the standard Kundur system [3], 2) the Kundur system as in [3], and 3) a large-scale model of the western interconnection, validated by duplicating the August 10, 1996 blackout event, to confirm our observations. The configuration of the Kundur test system from [3] is shown in Figure 8.

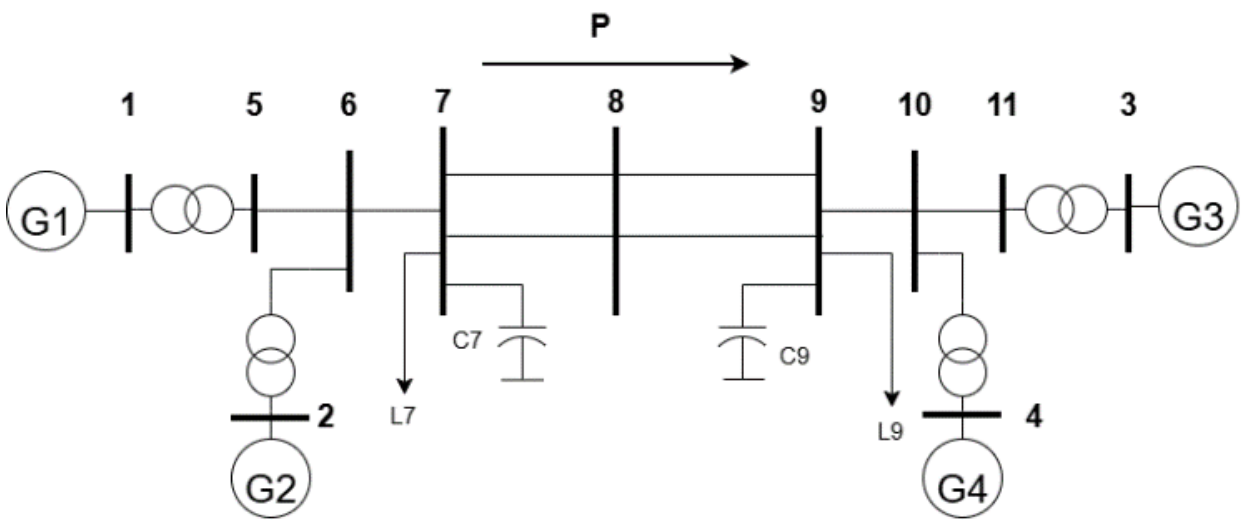

Figure 8: Classical Kundur test power system [3].

### III.1 Reduced Kundur system

As a first step, we simplified the standard Kundur system from [3] by removing generators 2 and 4 from Fig. 8. The resulting two-machine system provides a straightforward framework for examining the effect of transmission-line impedance on the damping of the inter-area oscillation mode. We performed a sensitivity analysis by varying the number of transmission lines and their lengths. The corresponding inter-area mode damping results are summarized in the tables below. In the base configuration, the system consists of two parallel tie-lines connecting bus 7 and bus 9 through bus 8, each 110 km long, and is highlighted in bold in all tables below. The **4-line** and **6-line** cases add parallel lines to the base model, resulting in a total of four and six parallel lines, respectively. All remaining system parameters are identical to those specified in Example 12.6 of the Kundur textbook [3].

In all cases, the per-unit damping constant '$K_d$' [3] was set to 0.5 for each generator in the model. To evaluate the impact of multiple transmission-line configurations, the shunt compensation levels were adjusted to 50 MVAR at Bus 7 and 100 MVAR at Bus 9, as reported in Table I whereas these shunt parameter values were kept unchanged from the textbook [3] for all subsequent cases and tables. The results presented in Tables I and II indicate an inverse relationship between transmission-line impedance and the damping of the inter-area mode: as the line impedance increases, the damping of the inter-area oscillation decreases. In Table I, the values shown in parentheses indicate the number of parallel transmission lines. Each parallel line is assumed to have identical series resistance, reactance, and charging-susceptance parameters.

TABLE I. SENSITIVITY ANALYSIS USING MUTILPLE LINES

| # of line | $R(\Omega)$ | $X_L(\Omega)$ | $B_C(\Omega^{-1})$ | *Freq (Hz)* | *Damping (%)* | *Tie-line flow (MW)* |
|---|---|---|---|---|---|---|
| 4 | 0.011(4) | 0.11(4) | 0.1925(4) | 0.85 | 7.81 | 377 |
| 6 | 0.011(6) | 0.11(6) | 0.1925(6) | 0.95 | 9 | 372 |

TABLE II. SENSITIVITY ANALYSIS USING LINE LENGTH

| Line length (km) | $R(\Omega)$ | $X_L(\Omega)$ | $B_C(\Omega^{-1})$ | *Freq (Hz)* | *Damping (%)* | *Tie-line flow (MW)* |
|---|---|---|---|---|---|---|
| 50 | 0.0025 | 0.025 | 0.175 | 0.96 | 7.85 | 371 |
| **110** | 0.0055 | 0.055 | 0.385 | 0.77 | 5.03 | 380 |
| 150 | 0.0075 | 0.075 | 0.525 | 0.63 | 1.22 | 388 |

## III.2 Standard Kundur system

After establishing the observed relationship in the simplified two-machine system, we performed the same analysis on the complete four-machine Kundur system. The results, presented in Tables III and IV, confirm the same overall trend: the damping of the inter-area mode decreases as the transmission-line impedance increases. In addition, we conducted a sensitivity analysis to investigate the influence of tie-line power transfer on inter-area mode damping. Tables V and VI present the results for high- and low-transfer operating conditions, respectively. These results demonstrate how changes in tie-line loading affect the inter-area oscillatory behavior and its associated damping. Under high-loading conditions, the damping of the inter-area mode changes significantly with variations in line impedance.

TABLE III. SENSITIVITY ANALYSIS USING MUTILPLE LINES

| # of line | $R(\Omega)$ | $X_L(\Omega)$ | $B_C(\Omega^{-1})$ | *Freq (Hz)* | *Damping (%)* | *Tie-line flow (MW)* |
|---|---|---|---|---|---|---|
| **2** | 0.011(2) | 0.11(2) | 0.1925(2) | 0.62 | 4.4 | 400 |
| 4 | 0.011(4) | 0.11(4) | 0.1925(4) | 0.81 | 6.51 | 389 |
| 6 | 0.011(6) | 0.11(6) | 0.1925(6) | 0.89 | 7.04 | 386 |

TABLE IV. SENSITIVITY ANALYSIS USING LINE LENGTH

| Line length (km) | $R(\Omega)$ | $X_L(\Omega)$ | $B_C(\Omega^{-1})$ | *Freq (Hz)* | *Damping (%)* | *Tie-line flow (MW)* |
|---|---|---|---|---|---|---|
| 50 | 0.0025 | 0.025 | 0.175 | 0.82 | 6.81 | 389 |
| **110** | 0.0055 | 0.055 | 0.385 | 0.62 | 4.4 | 400 |
| 150 | 0.0075 | 0.075 | 0.525 | 0.49 | 2.47 | 410 |

TABLE V. SENSITIVITY ANALYSIS USING LOW TRANSFER LINE

| # of line | $R(\Omega)$ | $X_L(\Omega)$ | $B_C(\Omega^{-1})$ | *Freq (Hz)* | *Damping (%)* | *Tie-line flow (MW)* |
|---|---|---|---|---|---|---|
| **2** | 0.011(2) | 0.11(2) | 0.1925(2) | 0.68 | 4.89 | 85 |
| 4 | 0.011(4) | 0.11(4) | 0.1925(4) | 0.83 | 5.34 | 84 |

TABLE VI. SENSITIVITY ANALYSIS USING HIGH TRANSFER LINES

| # of line | $R(\Omega)$ | $X_L(\Omega)$ | $B_C(\Omega^{-1})$ | *Freq (Hz)* | *Damping (%)* | *Tie-line flow (MW)* |
|---|---|---|---|---|---|---|
| **2** | 0.011(2) | 0.11(2) | 0.1925(2) | 0.54 | 2.39 | 517 |
| 4 | 0.011(4) | 0.11(4) | 0.1925(4) | 0.79 | 6.2 | 495 |

## III.3 1996 WECC -validated model

The 1996 WECC-validated model [1] was also used to test the same phenomenon, where the 0.25 Hz Northwest inter-area mode is the weakest inter-area mode in the region. The 1996 WECC system model comprises 6,564 buses, 1,005 generators, 6,375 transmission lines, and 3,208 loads. A diagram of the WECC system is shown in Figure 9. In this case, the tie-lines are the COI tie-lines, which consist of three lines: the first from Malin to Round Mountain listed as COI1, the second also from Malin to Round Mountain listed as COI2, and the third from Captain Jack to Olinda listed as COI3 in Table VII. We switched each line out one at a time to see whether the same relationship between line impedance and inter-area mode damping held. The results, listed in Table VII, show that while damping decreases in each case, switching line COI3 has a greater impact on inter-area mode damping than switching either of the other two lines.

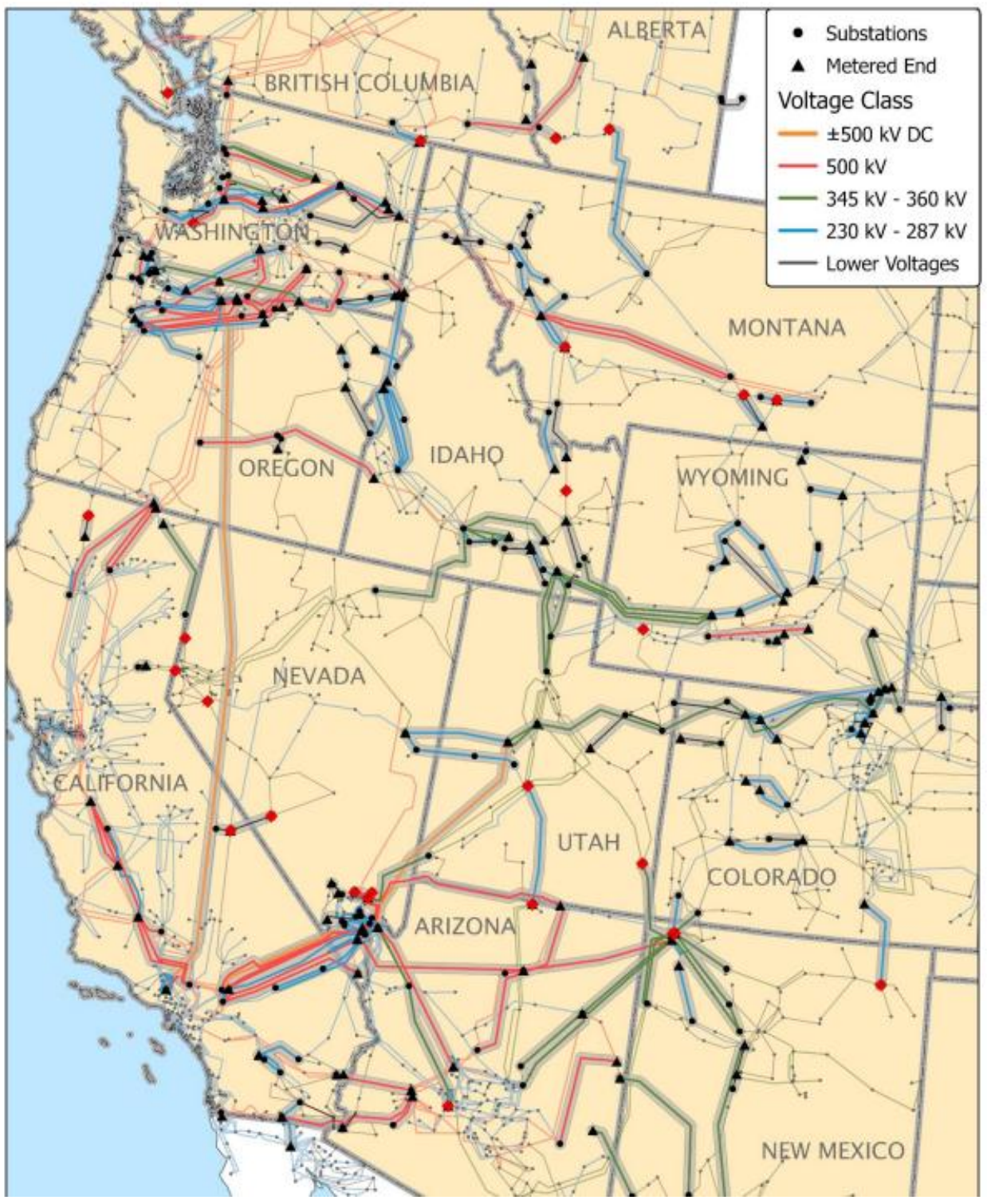

Figure 9: WECC system [9]

TABLE VII. SENSITIVITY ANALYSIS USING COI LINES IN WECC MODEL

| Line being switched | $R(\Omega)$ | $X_L(\Omega)$ | $B_C(\Omega^{-1})$ | *Freq (Hz)* | *Damping (%)* | *Tie-line flow (MW)* |
|---|---|---|---|---|---|---|
| **Base case** | | | | 0.27 | 4.79 | |
| COI1 | 0.00103 | 0.009 | 1.5804 | 0.26 | 2.05 | 4376 |
| COI2 | 0.00107 | 0.0089 | 1.527 | 0.26 | 1.98 | 4415 |
| COI3 | 0.0011 | 0.0123 | 2.9374 | 0.23 | -4.08 | 4986 |

## IV. Analysis of California ISO Real-time Models

California ISO serves as a reliability coordinator (RC) for Western Interconnection. RC West's Full Network Model (FNM) provides a detailed representation of the Western Interconnection and serves as the network foundation for dynamic simulations and stability assessments. The FNM represents system topology, generation and load conditions, transmission facilities, equipment parameters, and power transfers, providing the initial steady-state operating condition from which dynamic simulations are performed. The FNM is combined with the WECC-published seasonal dynamic models to simulate the system's response to disturbances such as transmission and generation outages, faults, and other significant system events. These simulations evaluate transient and voltage stability, frequency response, and oscillatory behavior, including identifying poorly damped inter-area modes and the system areas that may participate in them. The simulation results help identify potential stability-related reliability concerns and evaluate the operating limits and mitigation measures needed to maintain reliable operation of the RC West system.

These real-time dynamic models of the western interconnection are used to perform online dynamic stability studies periodically using a Dynamic Security Assessment (DSA) toolbox [7]. In this section, two archived large-scale models captured on April 2, 2026, at 5 pm and 10 pm, respectively, are used to study the sensitivity of transmission-line switching to damping of the 0.4 Hz North-South inter-area mode of the western system. The cases consist of 29,680 buses, 5,577 generators, and 15,027 loads.

### IV.1 April 2, 2026 5 pm model:

#### *A. No fault simulation:*

The COI MW transfer was at 1837 MW, and the Pacific DC Intertie (PDCI) flow was at 1010 MW for this case. California was importing power from the Pacific Northwest at that time. For the base case model, a 3-phase fault is applied at a COI tie-line bus for six cycles and cleared to excite the system dynamics. We analyze the resulting dynamic responses using ringdown analysis tools to estimate the modal composition of the signals [8]. Four algorithms, namely, Prony, Matrix Pencil (MP), Hankel Total Least Square (HTLS) method, and Eigenvalue Realization Algorithm (ERA) are used to verify the consistency of the results [8].

For the 5 pm base case, 193 generator frequency signals were saved from the simulations and were analyzed using the four algorithms, Prony, MP, HTLS, and ERA, to estimate the modal properties. For this base case, the analysis shows the dominant mode to be at 0.4 Hz with a damping ratio of 8.6%. Fig. 10a compares the modal analysis-based reconstructed signal (red) with the system response (black), and the reconstruction matches well. Fig. 10b shows the FFT of the same signal, which again shows that the reconstruction captures the nature of the 0.4 Hz mode very well in the frequency domain.

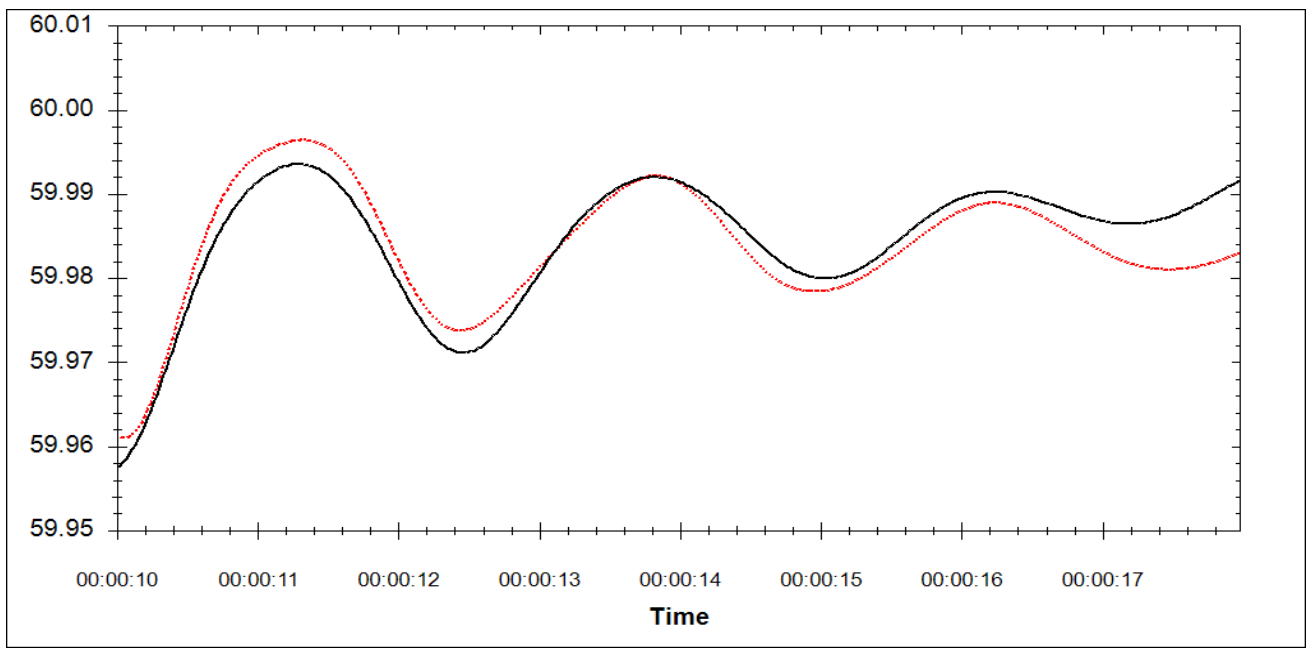

(a) Comparison of a simulated generator frequency plot (black) with the modal decomposition-based reconstruction (red)

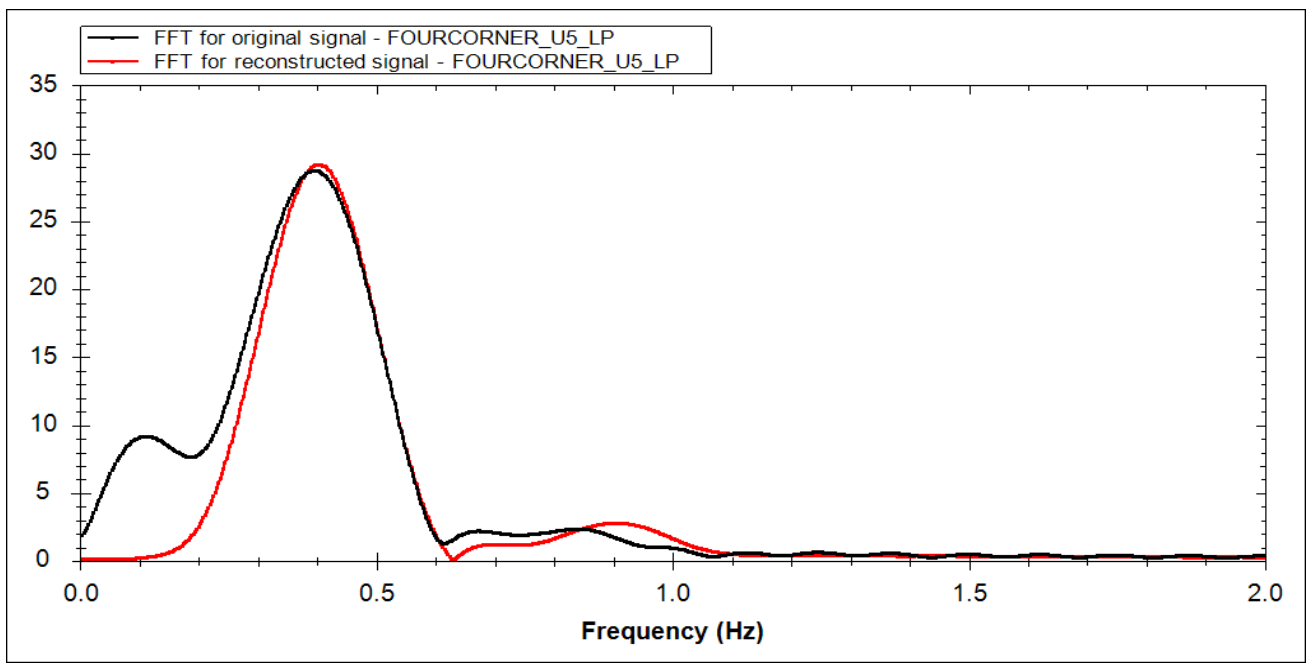

(b) Comparison of the FFT of the simulated generator frequency signal (black) with the FFT of the reconstructed signal (red)

Figure 10: Comparison of the modal analysis-based reconstructed signal with the simulated signal for the 5 pm base case model

*B. 500 kV double line outage:*

When two 500 kV lines in California are opened (Fig. 11), this severely stresses the system within California, and the damping of the 0.4 Hz mode decreases to 7.2%.

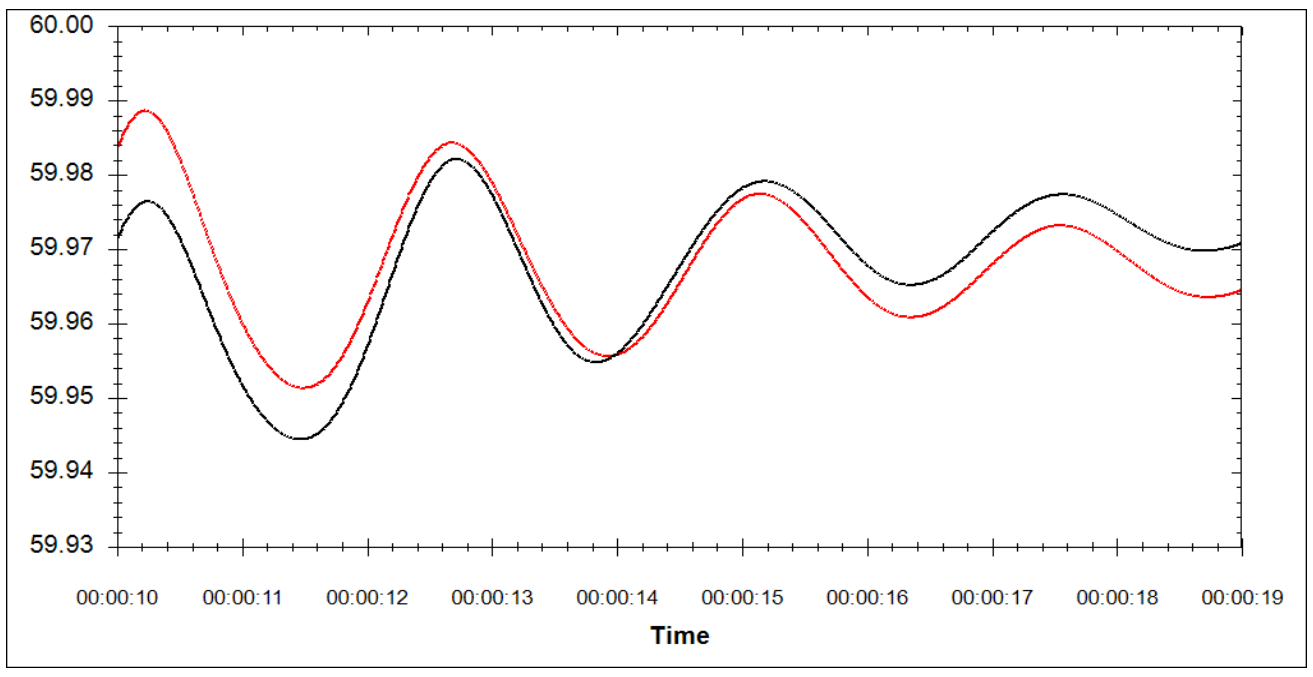

(a) Comparison of a simulated generator frequency plot (black) with the modal decomposition-based reconstruction (red)

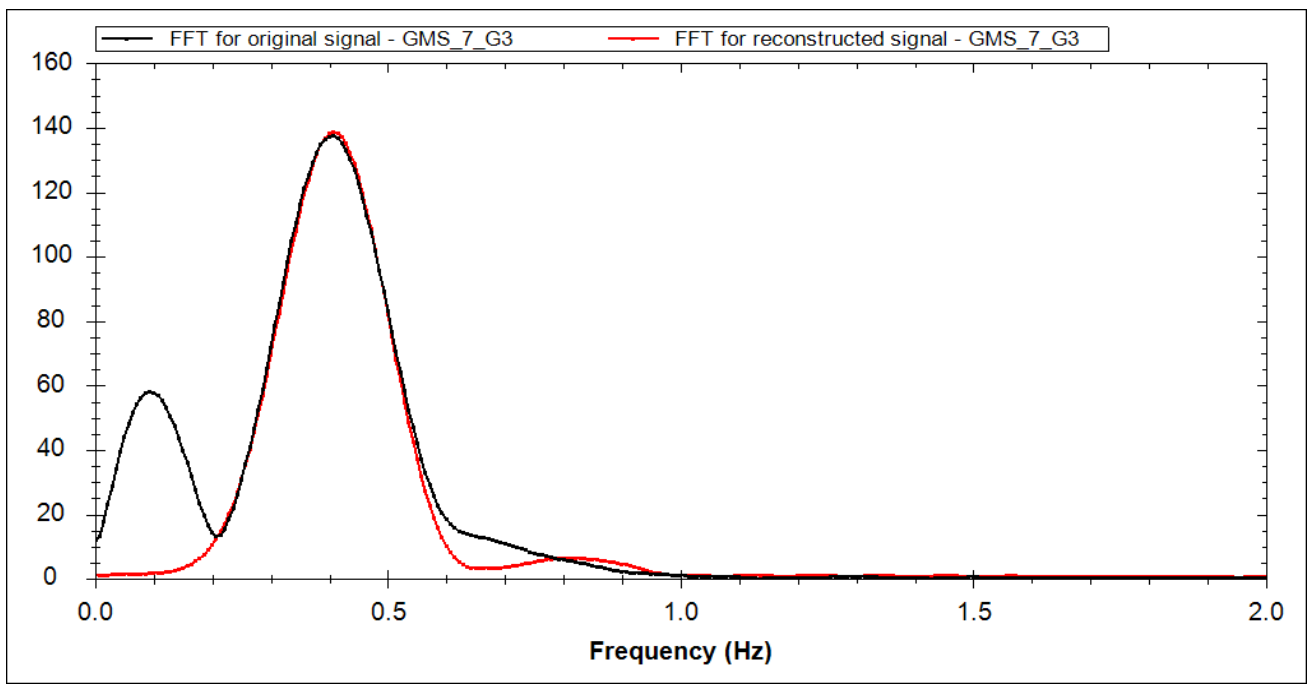

(b) Comparison of the FFT of the simulated generator frequency signal (black) with the FFT of the reconstructed signal (red)

Figure 11: Comparison of the modal analysis-based reconstructed signal with the simulated signal for the 5 pm model for the double contingency case

<u>IV.2 April 2, 2026 10 pm model:</u>

*A. No fault simulation:*

The COI MW transfer was at 4726 MW, and the Pacific DC Intertie (PDCI) flow was at 2505 MW for this case, and California was importing much higher power from the Pacific Northwest at that time compared to the 5 pm case in Section IV.1. Like before, a 3-phase fault is applied at a COI tie-line bus for six cycles and cleared to excite the system dynamics. The dominant mode frequency is 0.41 Hz, with a damping ratio estimated at 7.6%. The comparison of the modal decomposition-based reconstructed signal and the simulated signal, shown in Fig. 12, shows an excellent match. Compared with the 5 pm base case, the damping of the 0.4 Hz mode is slightly lower.

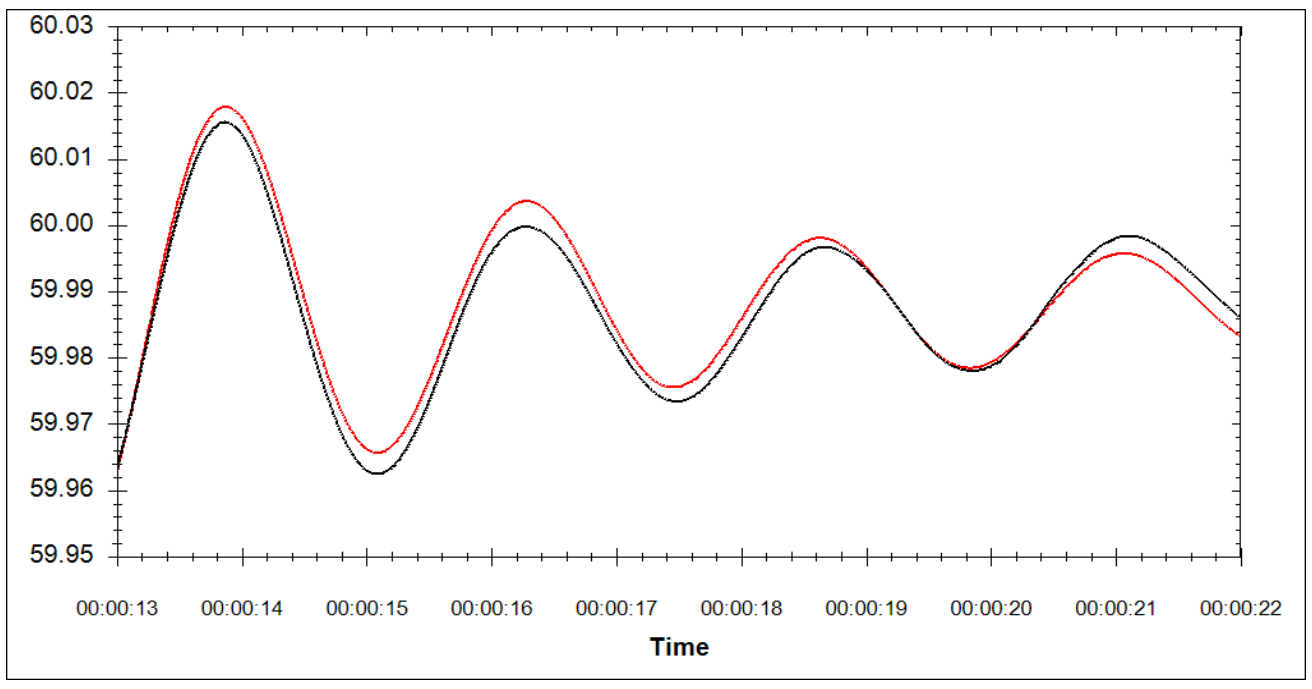

(a) Comparison of a simulated generator frequency plot (black) with the modal decomposition-based reconstruction (red)

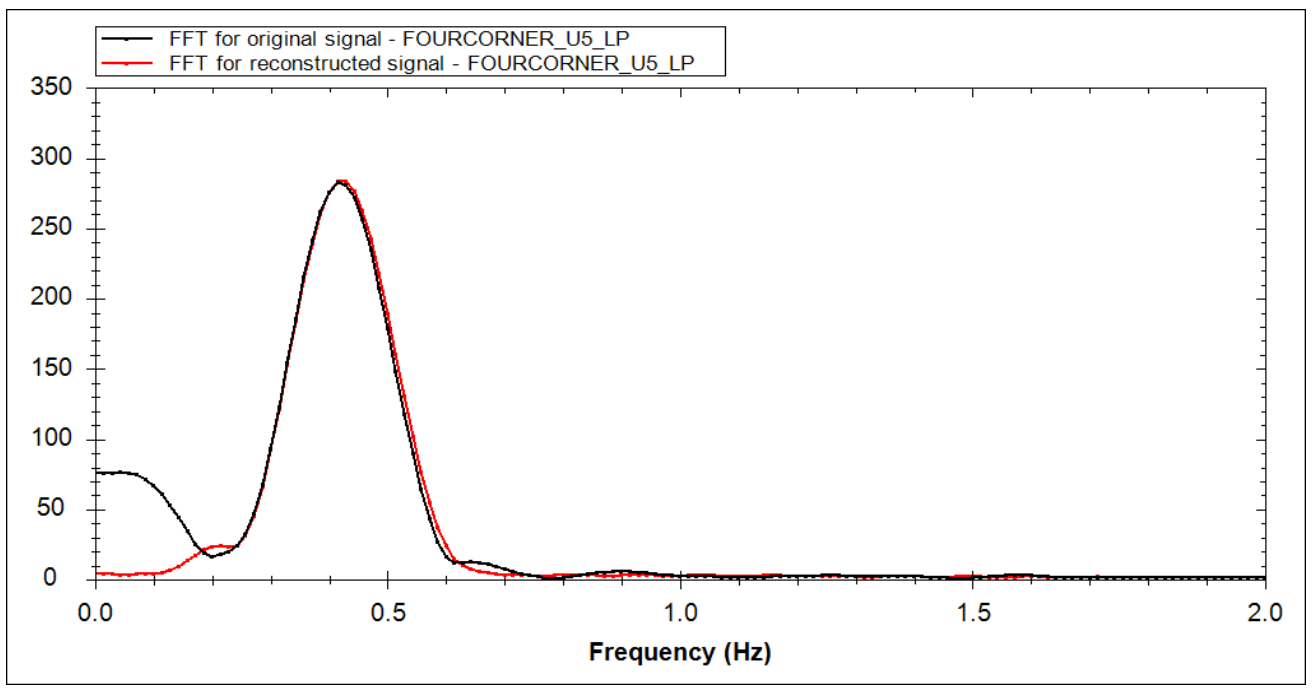

(b) Comparison of the FFT of the simulated generator frequency signal (black) with the FFT of the reconstructed signal (red)

Figure 12: Comparison of the modal analysis-based reconstructed signal with the simulated signal for the 10 pm base case model

### *B. 500 kV double line outage:*

When two California 500 kV lines are opened (Fig. 13), this reduces the 0.4 Hz mode damping. The mode frequency is estimated at 0.38 Hz and the damping ratio at 6.4%. We observe that the damping of the 0.4 Hz North-South Western Interconnection mode decreased for both the 5 pm and 10 pm cases, after the loss of two 500 kV lines. Conversely, the studies support industry practice of switching in key high-voltage AC transmission lines connecting the distant ends of the inter-area mode to improve its damping.

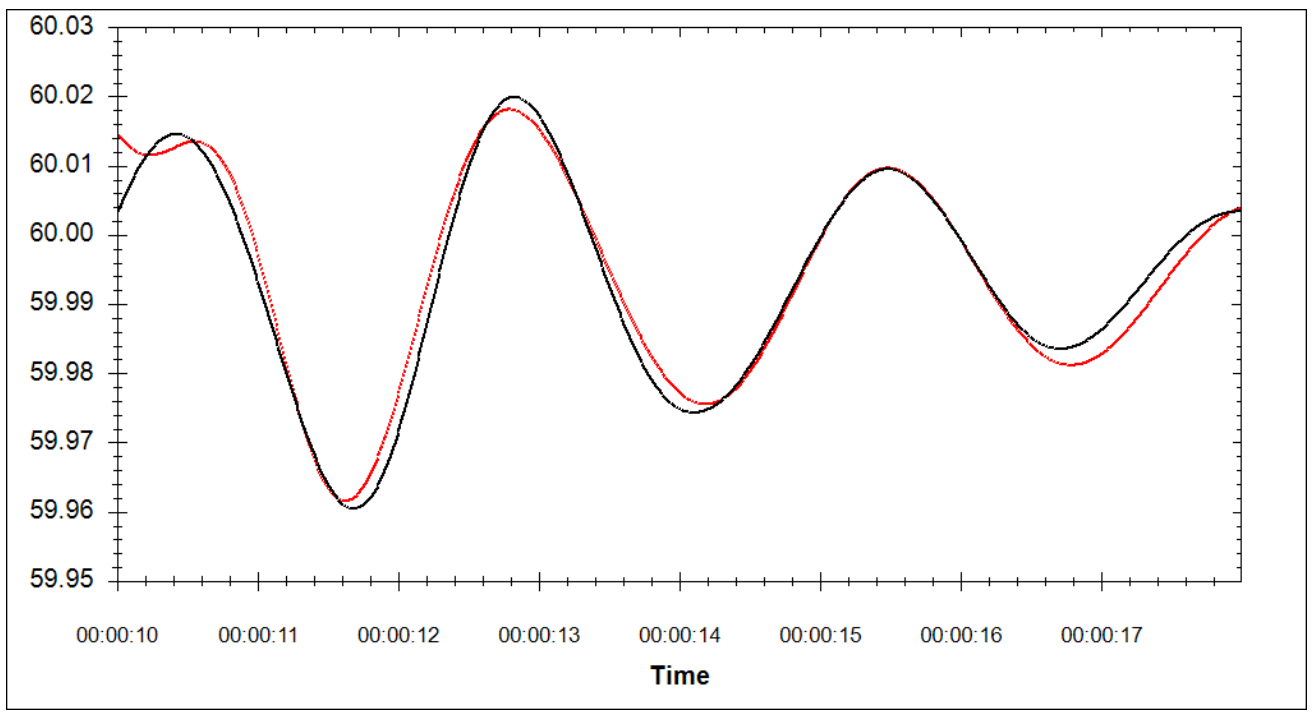

(a) Comparison of a simulated generator frequency plot (black) with the modal decomposition-based reconstruction (red)

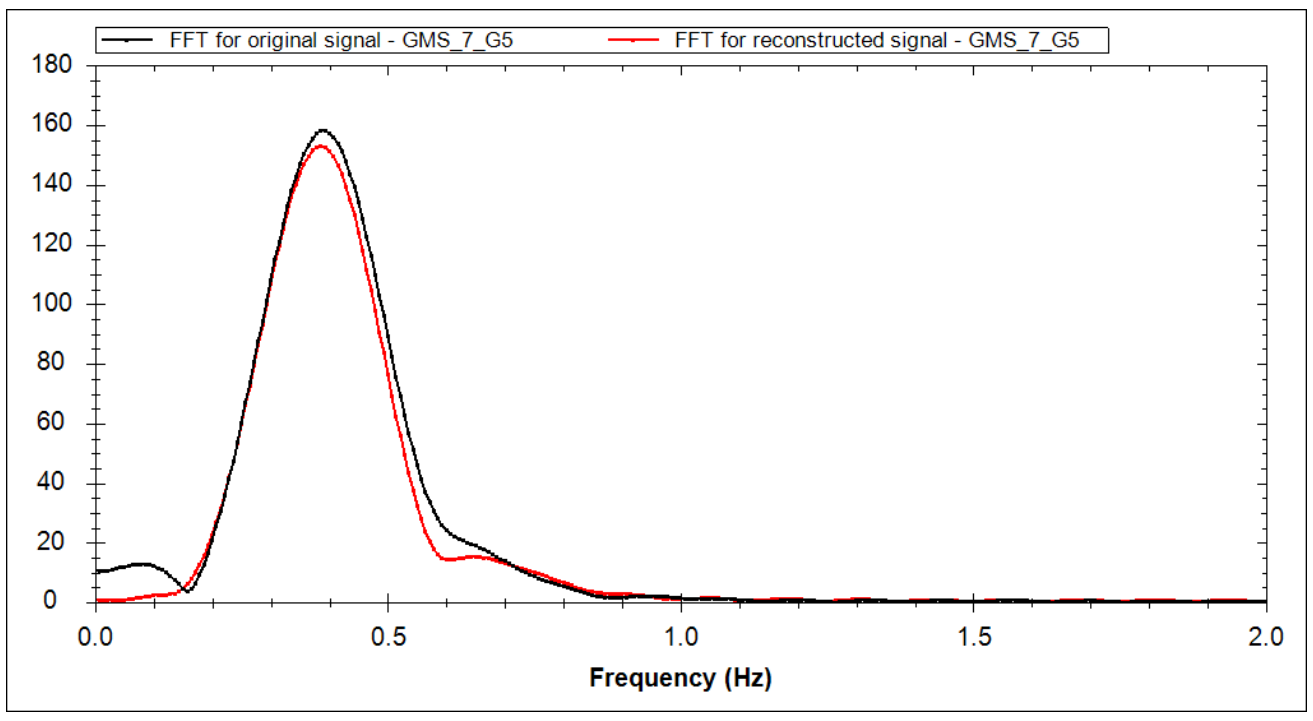

(b) Comparison of the FFT of the simulated generator frequency signal (black) with the FFT of the reconstructed signal (red)

Figure 13: Comparison of the modal analysis-based reconstructed signal with the simulated signal for the 10 pm model for the double contingency case

Simulations of the two archived Western system models from April 2, 2026, saved at 5 pm and 10 pm, both show that switching the two 500 kV lines has a similar effect on the damping of the 0.4 Hz Western interconnection mode. The mode frequency decreased slightly from 0.41 Hz at 5 pm to 0.38 Hz at 10 pm, while the damping ratio changed by about 1.4% and 1.2% at 5 pm and 10 pm, respectively, after the 500 kV double contingency. Conversely, the results show that switching in these two 500 kV transmission lines would have improved damping of the 0.4 Hz mode if the mode were stressed during operation and if the lines were out of service.

## V. Conclusion

This paper examines the relationship between transmission line strength, tie-line flows, and the damping of inter-area modes by analyzing past events, test-system models, and large-scale real-time system models. The results show that the loss of key transmission lines can significantly reduce the damping of critical inter-area modes. Conversely, inserting these lines during oscillation events can increase inter-area mode damping and improve system oscillatory stability. The impact of these actions is sensitive to system operating conditions. The damping effect may be greater under high MW transfer conditions and when the mode damping is lower. Further study is recommended to establish these observations theoretically.